\documentclass[footinbib,aps,pra,reprint,superscriptaddress]{revtex4-1}

\usepackage{amsmath,amssymb,braket,upgreek,bm,verbatim}
\usepackage{graphicx,tabularx} 
\usepackage[colorlinks=true,bookmarks=false,citecolor=blue,urlcolor=blue]{hyperref} %
\usepackage{xcolor}

\usepackage[absolute]{textpos}
\usepackage{xcolor}
\usepackage{multirow}

\newcolumntype{Y}{>{\centering\arraybackslash}X}

\begin{document}
\title{A Reconfigurable Quantum Local Area Network\\Over Deployed Fiber}

\author{Muneer Alshowkan}
\email{alshowkanm@ornl.gov}
\affiliation{Quantum Information Science Group, Oak Ridge National Laboratory, Oak Ridge, Tennessee 37831}
\author{Brian~P. Williams}
\author{Philip G. Evans}
\affiliation{Quantum Information Science Group, Oak Ridge National Laboratory, Oak Ridge, Tennessee 37831}
\author{Nageswara~S.~V. Rao}
\affiliation{Autonomous and Complex Systems Group, Oak Ridge National Laboratory, Oak Ridge, Tennessee 37831}
\author{Emma~M. Simmerman}
\affiliation{Department of Applied Physics, Stanford University, Stanford, California 94305, USA}
\author{Hsuan-Hao Lu}
\author{Navin~B. Lingaraju}
\author{Andrew~M. Weiner}
\affiliation{School of Electrical and Computer Engineering and Purdue Quantum Science and Engineering Institute, Purdue University, West Lafayette, Indiana 47907, USA}
\author{Claire~E. Marvinney}
\author{Yun-Yi Pai}
\author{Benjamin~J. Lawrie}
\affiliation{Materials Science and Technology Division, Oak Ridge National Laboratory, Oak Ridge, Tennessee 37831}

\author{Nicholas~A. Peters}
\author{Joseph~M. Lukens}
\email{lukensjm@ornl.gov}
\affiliation{Quantum Information Science Group, Oak Ridge National Laboratory, Oak Ridge, Tennessee 37831}

\date{\today}

\begin{abstract}
\label{Abstract}
Practical quantum networking architectures are crucial for scaling the connection of quantum resources.
Yet quantum network testbeds have thus far underutilized the full capabilities of modern lightwave communications, such as flexible-grid bandwidth allocation. In this work, we implement flex-grid entanglement distribution in a deployed network for the first time,  connecting nodes in three distinct campus buildings time-synchronized via the Global Positioning System (GPS). We quantify the quality of the distributed polarization entanglement via log-negativity, which offers a generic %
metric of link performance in entangled bits per second. %
After demonstrating successful entanglement distribution for two allocations of our eight dynamically reconfigurable channels, we demonstrate remote state preparation---the first realization on deployed fiber---showcasing one possible quantum protocol enabled by the distributed entanglement network.
Our results realize an advanced paradigm for managing entanglement resources in quantum networks of ever-increasing complexity and service demands.
\end{abstract}
\maketitle

\begin{textblock}{13.3}(1.4,15)
\noindent\fontsize{7}{7}\selectfont \textcolor{black!30}{This manuscript has been co-authored by UT-Battelle, LLC, under contract DE-AC05-00OR22725 with the US Department of Energy (DOE). The US government retains and the publisher, by accepting the article for publication, acknowledges that the US government retains a nonexclusive, paid-up, irrevocable, worldwide license to publish or reproduce the published form of this manuscript, or allow others to do so, for US government purposes. DOE will provide public access to these results of federally sponsored research in accordance with the DOE Public Access Plan (http://energy.gov/downloads/doe-public-access-plan).}
\end{textblock}

\section{Introduction}
\label{Introduction}
Quantum communications networks play a key role in advancing quantum information science (QIS).  Some examples where quantum networks are crucial include distributed computing~\cite{Cirac1999}, enhanced sensing~\cite{Bollinger1996, Giovannetti2004, Giovannetti2011}, secure communications~\cite{Bennett2014, Gisin2007},  blind computing~\cite{Broadbent2009, Barz2012} and the highly anticipated quantum internet~\cite{Kimble2008, Wehner2018}.
In the context of fundamental scientific discovery, quantum networks have potential to improve the sensitivity of astronomical interferometry~\cite{Gottesman2012, Khabiboulline2019}, and networks attaining entanglement-enhanced clock synchronization~\cite{Jozsa2000, Komar2014, Komar2016} should impact a variety of detection techniques based on distributed atomic clocks, including those being explored in dark matter searches~\cite{Roberts2017, Wcislo2018}.

Practical quantum key distribution (QKD)~\cite{Scarani2009, Xu2020}---arguably the most mature quantum application---is designed for  two-node links; in order to scale beyond two nodes for quantum digital signatures~\cite{Gottesman2001, Clarke2012, Dunjko2014} or secret sharing~\cite{Hillery1999, Karlsson1999, Tittel2001, Chen2005, Schmid2005, Grice2015, Williams2019}, fully connected quantum network architectures are desirable.

\label{Literature Survey}
Fundamentally, any architecture should support entanglement between distant parties on-demand as a core capability, %
ideally in an efficient and agile manner.
Theoretical approaches to extend distance based on quantum repeaters~\cite{Briegel1998} are promising; however, the current technology is still in very early stages. %
The current implementations of quantum networks at the logical level can be classified as point-to-point, trusted-node, point-to-multipoint, and fully connected. %
The simple point-to-point link is found in typical QKD implementations between two remote parties, traditionally known as Alice and Bob~\cite{Scarani2009}.
The trusted-node network~\cite{Elliott2002, Peev2009, Stucki2011, Sasaki2011, Chen2010, Wang2014, Mao2018, Dynes2019, Evans2019} consists of multiple point-to-point links in a partial mesh fashion, where optical links terminate in trusted nodes designed for a given Alice-Bob pair. 
Communications between distant nodes are enabled by intermediate nodes in a hop-by-hop paradigm~\cite{Elliott2002, Peev2009}, so that end-to-end security requires that all intermediary nodes be trusted.
In the simplest case of point-to-multipoint networks, a passive beamsplitter can be used to enable a node to communicate with one of any available nodes at random~\cite{Townsend1997}. 
Alternatively, leveraging broadband frequency and polarization hyperentangled photons~\cite{Lim2008}, dedicated entanglement can be established between any pair of users by assigning frequency-correlated wavelength channels. Fully connected quantum networks have recently been realized in this paradigm using nested dense wavelength division multiplexers (DWDMs)~\cite{Wengerowsky2018, Joshi2020}. Reconfigurable quantum links can be obtained by combining such DWDMs with transparent spatial switches~\cite{Toliver2003, Peters2009, Chapuran2009, Herbauts2013, Laudenbach2020}, although the individual channel spacings and bandwidths remain fixed. By permitting flexible grid definitions in addition to spatial switching, a wavelength-selective switch (WSS) offers further improvements, and has recently been leveraged in fully connected quantum network designs utilizing adaptive resource provisioning~\cite{Lingaraju2020a, Appas2021}.

\begin{figure*}[tb!]
	\centering
	\includegraphics[width=\textwidth]{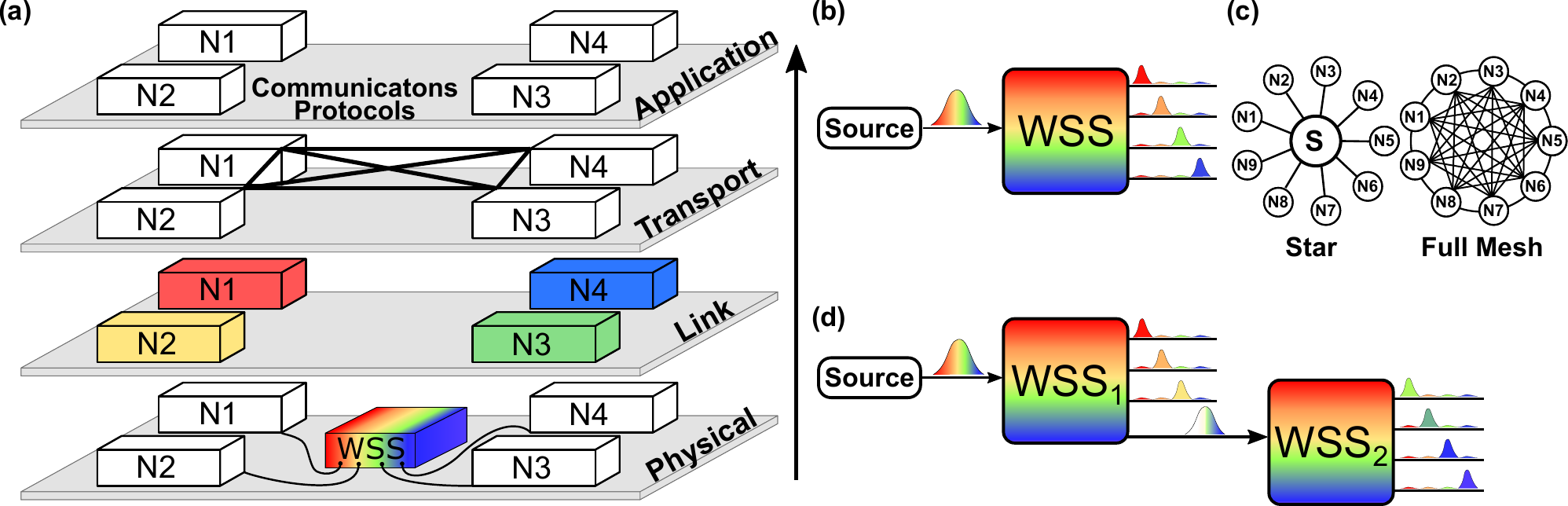}
	\caption{(a) QLAN communication layers and services. The physical layer includes the optical components where the photons travel and are manipulated, the link layer slices the spectrum and routes each slice to a particular user, the transport layer is the quantum-correlated network where a pair of users shares entanglement, and the application layer uses the entangled pairs to perform a service. (b) Single WSS configuration: the input spectrum is sliced and routed to the output ports. (c) Comparison of physical (star) and logical (fully connected mesh) topologies in network design. (d) Nested WSS where a portion of the spectrum is routed from the first to the second WSS, thereby expanding the number of nodes connected by the entanglement source.}
	\label{fig2}
\end{figure*}

Although showing great promise, %
results for fully connected quantum entanglement networks with dense wavelength allocation have  been based on either loop-back fibers~\cite{Joshi2020} or a tabletop configuration~\cite{Wengerowsky2018,Lingaraju2020a,Appas2021} where all detection events occur in the same physical site.
This simplification of time synchronization and data management procedures does not reflect the ultimate networking objective of nonlocal spatially distributed entanglement. Practically useful quantum networks must consist of spatially separated nodes with independent, heterogeneous quantum resources (stationary qubits, detectors, photon sources, etc.) synchronized to a common clock. %
The network architecture must be compatible with other networking topologies to enable forming larger networks. A quantum network will also need classical network capabilities in the form of a control plane for management and a parallel  data plane for classical communications between the network nodes. %

\label{Goal of this paper}
Here, we address these needs in the first dynamic, fully connected quantum local area network (QLAN) for entanglement distribution over deployed fiber. Combining both adaptive bandwidth provisioning and simultaneous distributed remote detection with off-the-shelf control systems, the three-building campus network has eight independent entanglement channels in the lowest-loss telecommunications transmission window that are dynamically and seamlessly remotely reconfigurable, thus allowing various configurations and bandwidth allocations without the need to add  or remove any components in the setup.  %
By measuring link throughput in terms of entangled bits per second (ebits/s), we validate the quality of each logical connection and show how different provisioning scenarios modify these ``on demand'' rates. Finally, to concretely demonstrate that the network supports quantum protocols, we realize remote state preparation (RSP) over all links, which  to our knowledge is the first implementation of this quantum communications protocol in any deployed network. Overall, our layered network design and use of flex-grid lightwave technology furnish %
a promising concept and path toward more general quantum networks, including those supporting quantum computers and quantum sensors, ultimately establishing a framework on which a future quantum internet can be architected.

\label{Significance of the work}

\section{Setup}
\label{Setup}

\subsection{Network Architecture}
\label{Network Architecture}
\label{Introduction to TCP/IP stack and how it relates to our network}
The QLAN can be described in terms of functional layers shown in Fig.~\ref{fig2}(a) in a manner analogous to the Transmission Control Protocol/Internet Protocol (TCP/IP) stack~\cite{Cerf1974, braden1989}.
In general, this logical structure encompasses the physical medium, routing, protocols, 
and applications that describe the behavior of each node in the network.
In this way, we seek to bring the quantum network architecture closer to practice using the layered model. 
In some ways, the quantum network layers may closely mirror their classical counterparts; e.g., the point-to-point links at the link layer are quite similar to optical channels provisioned over conventional network backbones using add/drop DWDM devices.
On the other hand, quantum network architectures diverge from classical networks in crucial ways. For example, the no-cloning theorem prohibits making a perfect copy of an arbitrary unknown quantum state~\cite{Wootters1982}, thus preventing the use of conventional detection and retransmission techniques in optical routing.
For this reason, quantum layers may not offer the same functionalities found in their classical counterparts (e.g., error detection and correction as in the TCP/IP stack link layer)~\cite{Wehner2018}, and may require fundamental refinements as the technology advances.

\label{Decription of the layers in our network}
In the QLAN, the physical layer corresponds to the physical components where the encoded photons are transmitted, received, and manipulated over the communications medium (optical fiber in this case). %
Every user in the network is connected to the WSS via the deployed fiber [Fig.~\ref{fig2}(b)]; thus, the physical connections of this network correspond to the star topology option depicted in Fig.~\ref{fig2}(c).
In the link layer, the WSS partitions the received bandwidth and dynamically routes the correlated spectral slices to particular output ports. %
In terms of  entanglement distribution, the transport layer can be viewed as a logical fully connected mesh topology;
each link forms a private point-to-point connection between two nodes as in the ``Full Mesh'' option in Fig.~\ref{fig2}(c). Because of the nonlocal nature of quantum entanglement, private \emph{logical} connections appear even without a corresponding direct \emph{physical} link: in other words, a star physical topology---central source connected to all $N$ nodes---is able to support a logical fully connected mesh, with entanglement connections between all $N(N-1)/2$ user pairs.
The application layer utilizes these logical connections to provide a quantum service between the connected users (e.g., QKD, teleportation, RSP).
Building on similar ideas proposed in the design of quantum access networks~\cite{Ciurana2015, Alshowkan2016}, our architecture can be scaled by nesting multiple WSSs to create much larger networks %
as shown in Fig.~\ref{fig2}(d). 
For example, the first WSS could %
send a section of the spectrum to the second WSS, creating a %
fully connected network where entangled links are formed between nodes in both WSSs. %
Note that in the QLAN stack here, we do not include a layer analogous to the internet %
layer found in the classical TCP/IP conceptual model~\cite{braden1989}. 
Located between link and transport, the internet layer provides internetworking to connect multiple networks together%
---a capability not yet required in our single-network testbed.

At this early stage of development, any proposed quantum network stack is admittedly tentative, subject to refinements and modifications as quantum networks continue to mature. %
Yet by proposing such a stack here and identifying our QLAN's capabilities with relevant classical analogues, we are able to hint at the level of abstraction possible even with relatively simple quantum networks, in turn demonstrating a promising path forward for managing quantum resources in a scalable fashion.

\subsection{Source Description}
\label{Source Description}
Our QLAN utilizes an entangled photon source based on a 10~mm-long, periodically poled lithium niobate (PPLN) waveguide (HC Photonics), engineered for type-II spontaneous parametric down-conversion (SPDC). Pumping the PPLN waveguide with a continuous-wave laser ($\lambda= 779.4$ nm for degenerate SPDC) followed by temporal walk-off compensation and spectral separation~\cite{Kaiser2012,Lingaraju2020a}, we can prepare signal-idler photon pairs which ideally are spectrally and polarization entangled, as described by the state
\begin{equation}
\begin{split}
\ket{\psi} \propto & \int d\omega \,\Phi(\omega) \Big[\hat{a}^\dagger_H(\omega) \hat{a}^\dagger_V(2\omega_0-\omega) \\ & + e^{i\phi} \hat{a}^\dagger_V(\omega) \hat{a}^\dagger_H(2\omega_0-\omega) \Big] \ket{\mathrm{vac}}.
\end{split}
\end{equation}
Here $\Phi(\omega)$ is the crystal phase-matching function with full-width at half-maximum bandwidth of $\sim$2.5~nm (310~GHz), $\hat{a}^\dagger_{H(V)}(\omega)$ generates a photon at frequency $\omega$ with horizontal (vertical) polarization, and $\omega_0$ is half the pump frequency ($\omega_0/2\pi=192.3125$~THz). The residual phase $\phi$ is compensated by the method described in Sec.~\ref{depNet}, leaving the ideal Bell state $\ket{\Psi^+} = \frac{1}{\sqrt{2}}(\ket{HV}+\ket{VH})$ for every pair of energy-matched frequencies. Further details regarding the preparation of the photon source can be found in Ref.~\cite{Lingaraju2020a}.

This state is then sectioned into bands that are distributed to network users. We utilize a WSS to define 8 pairs of frequency-correlated channels. Each channel $n$ has a width of $\Delta\omega/2\pi=25$~GHz and aligns to the International Telecommunication Union (ITU) grid (ITU-T Rec. G.694.1) as listed in Table.~\ref{table2}, centered at $\omega_n = \omega_0 \pm \Delta\omega(n-\frac{1}{2})$ for the signal (idler). These channels occupy a $\sim$3~nm bandwidth in the C-band (1557.3--1560.5~nm). %

\begin{table}[tb!]
	\centering
	\caption{DWDM ITU $25$~GHz grid C-band channel numbers of the signal and the idler photons. The pair of channels in the same row are entangled. Fidelities are with respect to the ideal polarization Bell state $\ket{\Psi^+}$, and both photons are measured locally at the source prior to transmission.}
	\label{table2}
	\begin{ruledtabular}
	\begin{tabular}{c|cc|cc|c}
			\multirow{2}{*}{\textbf{Ch.}} & \multicolumn{2}{c|}{\textbf{Signal}} & \multicolumn{2}{c|}{\textbf{Idler}} &
			\multirow{2}{*}{\textbf{Fidelity}} \\ %
		& \textbf{ITU} & \textbf{THz}  & \textbf{ITU} & \textbf{THz} &  \\ \hline
			1 & 23.25  & 192.325  & 23.00 & 192.300 & $0.952 \pm 0.007$ \\
			2 & 23.50  & 192.350  & 22.75 & 192.275 & $0.948 \pm 0.007$ \\
			3 & 23.75  & 192.375  & 22.50 & 192.250 & $0.942 \pm 0.008$ \\
			4 & 24.00  & 192.400  & 22.25 & 192.225 & $0.935 \pm 0.007$ \\
			5 & 24.25  & 192.425  & 22.00 & 192.200 & $0.944 \pm 0.006$ \\
			6 & 24.50  & 192.450  & 21.75 & 192.175 & $0.949 \pm 0.006$ \\ 
			7 & 24.75  & 192.475  & 21.50 & 192.150 & $0.943 \pm 0.007$ \\ 
			8 & 25.00  & 192.500  & 21.25 & 192.125 & $0.947 \pm 0.007$ \\
	\end{tabular}
	\end{ruledtabular}
\end{table}

\begin{figure}[tb!]
	\centering
	\includegraphics[scale=0.42]{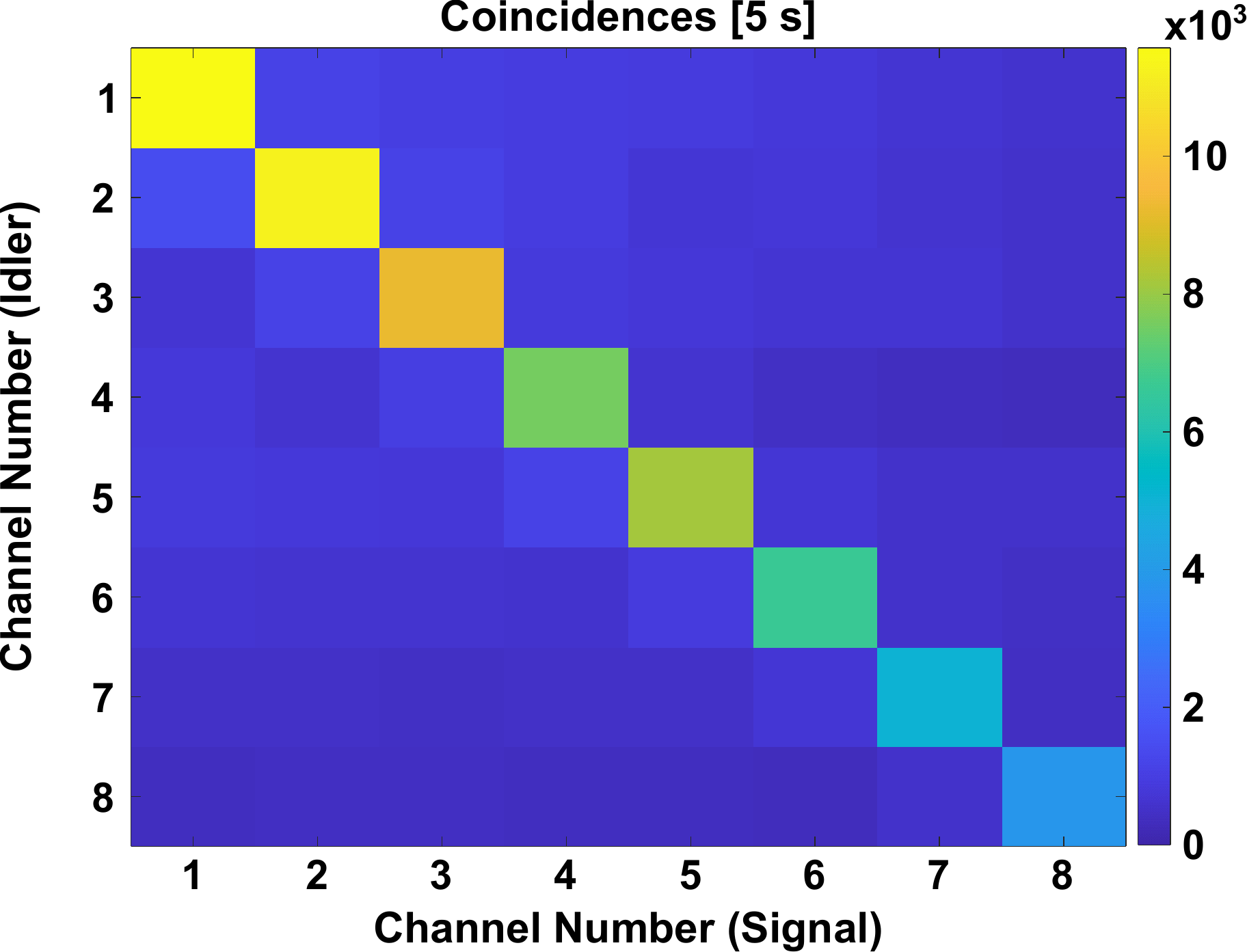}
	\caption{
	Joint spectral intensity of the source measured locally with SNSPDs and plotted as a function of the signal and idler channel numbers. Each point is obtained with a 5~ns coincidence window and 5~s integration time.
	}
	\label{jsi}
\end{figure}

\label{Characterization}
We first characterize the spectral correlations and polarization entanglement of the source with two co-located receivers. For joint spectral intensity (JSI) characterization, we program the WSS to route one signal channel and one idler channel to two superconducting nanowire single-photon detectors (SNSPDs) for coincidence measurement. Figure~\ref{jsi} shows the obtained JSI with a 5~ns coincidence window and 5~s integration time per measurement (at pump power $\sim$17~mW). The channel numbers for the 25~GHz-wide signal and idler bins correspond to the ITU mapping in Table~\ref{table2}. Strong correlations are observed for energy-matched channels, with small crosstalk in the lower off-diagonal. Overall, the coincidence-to-accidental ratio (average of matched bins divided by average of mismatched bins) is 11.4.

To characterize the quality of polarization entanglement for these eight channel pairs, we measure two-photon correlations in the $H/V$ and $D/A$ bases, defining $\ket{D}=\frac{1}{\sqrt{2}}(\ket{H}+\ket{V})$ and $\ket{A}=\frac{1}{\sqrt{2}}(\ket{H}-\ket{V})$. We employ the Bayesian tomography method of Ref.~\cite{Lukens2020b} for quantum state reconstruction such that, though two sets of measurements (both in $H/V$ and $D/A$ bases) are not tomographically complete, we can still obtain meaningful state estimates with low error bars due to the highly correlated nature of our input state. We confirm high fidelity with respect to $\ket{\Psi^+}$ for each individual biphoton channel as shown in Table~\ref{table2}.%

\label{What need to be solved}
\subsection{Time Synchronization}
\label{Components of each node}
As every node in the network is spatially distributed, independent receivers are required to evaluate network performance and perform quantum communications tasks. For accurate photon coincidence counting, all time-to-digital converters (TDCs) in the network must be synchronized to a common clock. Such distributed timing synchronization presents a critical technical challenge in extending tabletop quantum networking experiments to the field. And the timing requirements in quantum communications are frequently much tighter than for classical LANs. For example, the precision time protocol (PTP) is a simple and ubiquitous standard for synchronizing devices over ethernet, yet it is designed to attain only sub-$\upmu$s jitter~\cite{IEEE1588}, whereas sub-ns precision is desired in many photon counting quantum networking experiments. As a major improvement on PTP, White Rabbit~\cite{Linpinski2011} leverages synchronous ethernet to attain ps-level precision. Accordingly, White Rabbit appears particularly promising for future quantum networks, but requires protocol-compatible ethernet infrastructure. Alternatively, as lower-level, fully optical approaches, quantum communications experiments have utilized optical sync pulses %
or direct tracking of either pulsed quantum signals~\cite{Hughes2005, Chapuran2009}
or photon coincidence peaks~\cite{Peloso2009, Krenn2015, Steinlechner2017, Shi2020} to compensate for the drift in independent clocks.

\begin{figure}[tb!]
	\centering
	\includegraphics[scale=0.4]{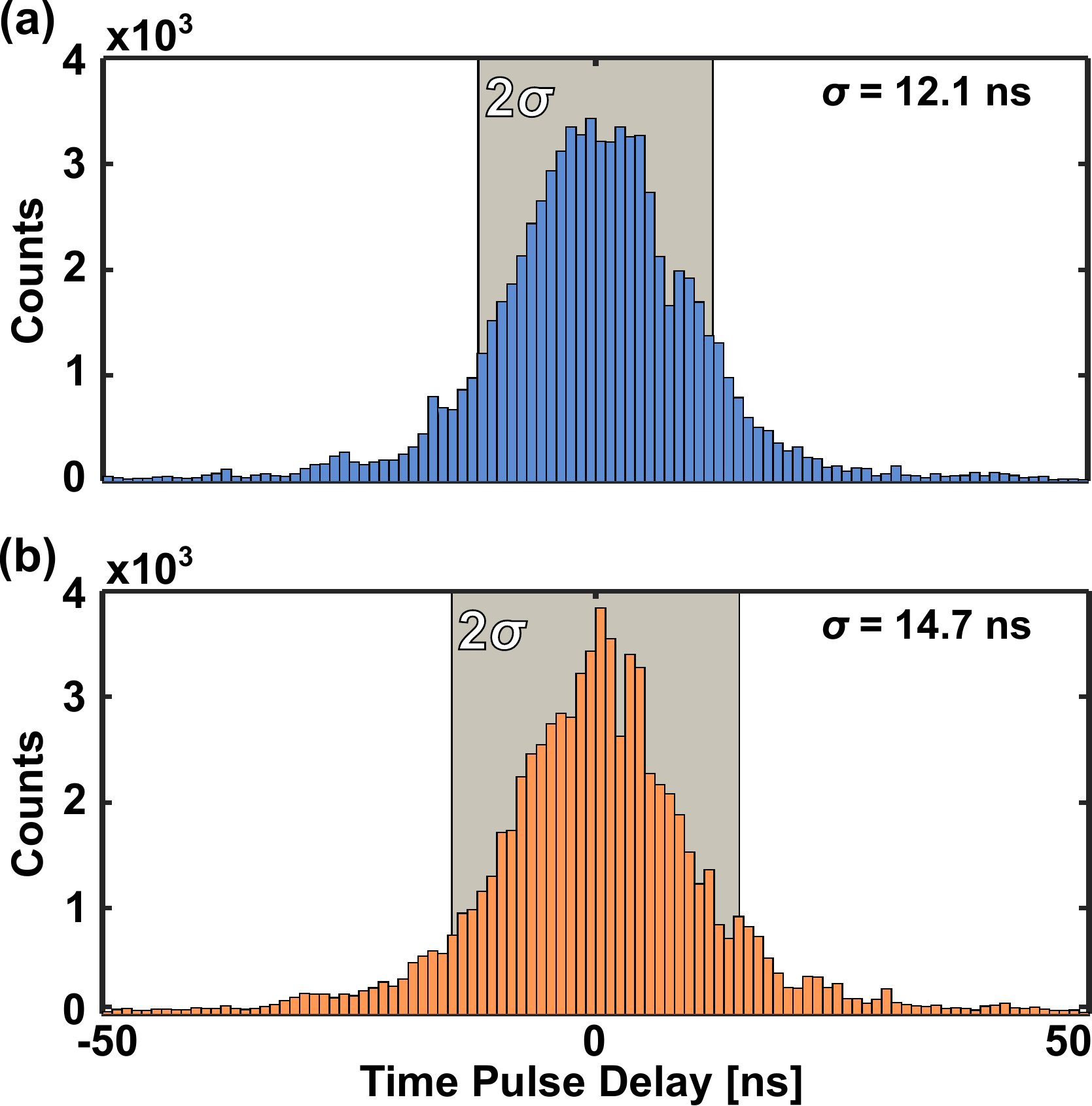}
	\caption{Histogram of relative delays between time pulses from two GPS receivers. (a) Alice and Bob. (b) Charlie and Alice.}
	\label{jitter}
\end{figure}

\label{GPS and Jitter}
In our QLAN, we selected GPS-based synchronization for its simplicity, cost-effectiveness, and availability (a GPS antenna was already available in one lab). From the GPS signal, both time pulses---at a rate of one pulse per second (PPS)---and a 10~MHz clock can be derived. Before deployment on our network, we characterized the jitter between time pulses produced by pairs of independent receivers (Trimble Thunderbolot E). Histograms of the relative delays recorded over several hours using a time interval counter (Stanford Research Systems SR620) appear in Fig.~\ref{jitter}, for two pairs of the three devices. The distributions of relative delays have standard deviations of 12.1~ns for the devices of Alice and Bob [Fig~\ref{jitter}(a)] and 14.7~ns for Charlie and Alice [Fig.~\ref{jitter}(b)]. This suggests coincidence windows up to $\sim$30~ns would be reasonable in our network; in practice, we found a window of 10~ns to offer a reasonable balance between counting photon pairs and reducing the noise from accidentals. At these jitters, GPS offers significantly higher precision than PTP, though upgrading to a system reaching sub-ns levels would be an important improvement moving forward.

\section{Implementation}
\label{Implementation}
\label{outline of the network}
\subsection{Deployed Network}
\label{depNet}
\begin{figure*}[t!]
	\centering
	\includegraphics[width=\textwidth]{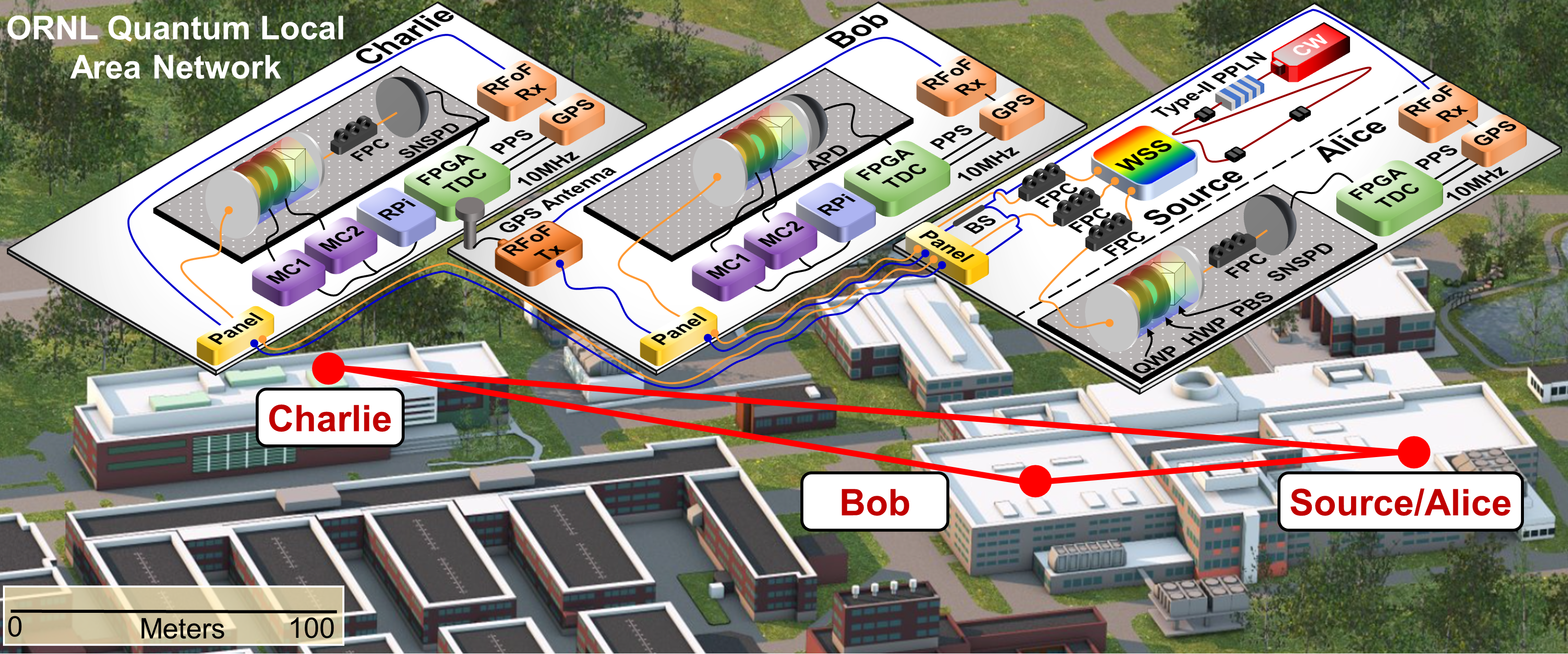}
	\caption{Map of QLAN on ORNL campus. The receiver configurations at each node are shown as insets. APD: avalanche photodiode. CW: continuous-wave laser. FPC: fiber polarization controller. FPGA: field-programmable gate array. GPS: Global Positioning System. HWP: half-wave plate. MC: motion controller. Panel: fiber-optic patch panel. BS: beamsplitter. PBS: polarizing beamsplitter. PPLN: periodically poled lithium niobate. PPS: pulse per second. Source: entanglement source. QWP: quarter-wave plate.  RFoF Rx: RF over fiber receiver. RFoF Tx: RF over fiber transmitter. RPi: Raspberry Pi microprocessor board (to control MCs). SNSPD: superconducting nanowire single-photon detector. WSS: wavelength-selective switch.}
	\label{setup_map}
\end{figure*}

The experiment was performed in three buildings on the Oak Ridge National Laboratory (ORNL) campus, as depicted in Fig.~\ref{setup_map}. The source and Alice are co-located in the same lab while Bob and Charlie reside in separate buildings, connected to the source through total fiber path lengths of approximately 250~m and 1200~m, respectively. The source of the polarization-entangled photons and each user in the network are connected to a WSS by a single fiber, where the outputs %
pass through a manual fiber-based polarization controller (FPC) to compensate for the polarization rotation along the fiber path. 
The output of the FPC goes directly to the polarization analyzer for Alice and to the fiber patch panel for Bob and Charlie.
From the patch panel, the fibers traverse multiple communication rooms before reaching the panels at Bob and Charlie.
The fiber from the patch panel in each location is then routed to the polarization analyzer. %
The output of Bob's polarization analyzer is sent directly to an InGaAs avalanche photodiode (APD), whereas for Alice and Charlie, the outputs are directed to SNSPDs preceded by FPCs to maximize detection efficiency.

While each user could employ a local antenna, it was simpler in our case to share the GPS signal from the antenna at Bob's location with the other nodes via RF over fiber (RFoF).
The RF GPS signal at Bob is connected to a commercial RFoF transmitter that outputs an optical signal. This is subsequently split and sent over separate strands of the deployed fiber to every node's RFoF receiver, which then outputs the original RF GPS signal.
The GPS receivers at each network node synchronize an FPGA-based TDC with the 10~MHz and 1~Hz (PPS) clocks; we bin each detection event to 5~ns resolution using the FPGA's internal 200~MHz clock, tracking the number of 10~MHz and 1~Hz cycles to assign a global network timestamp.
Note that our approach offers a general and scalable framework for increasing the network size, as bringing an additional node online requires a single-photon detector and relatively inexpensive off-the-shelf electronic components.

\label{Need for Control Plane}
For data transfer and instrument control between the nodes, we set up a %
classical network managed by a control plane for routing over the campus networking infrastructure.
Events recorded by each TDC are shared over the control plane for data analysis and coincidence counting.
\label{Need for classical network}
This dedicated network allows timely and accurate data transfer for analysis and monitoring. It is crucial to have a classical network with bandwidth commensurate with the quantum data gathered. In our experiments, each time stamp is a 32-bit record, and events occur at rates up to $\sim$10$^5$ s$^{-1}$ at the SNSPD nodes, corresponding to an average data rate of $\sim$3~Mb/s---comfortably below the $\sim$1~Gb/s supported by our control plane.

With quantum information encoded in polarization in the QLAN, 
each node holds a polarization analyzer, a single-photon detector, and a TDC.
Components of the analyzer include a quarter-wave plate (QWP), half-wave plate (HWP), and polarizing beamsplitter (PBS). %
The angles of the waveplates at Bob's and Charlie's locations are set remotely with motion controllers, and manually at Alice's location.
Alice and Charlie possess SNSPDs with efficiency $\eta \geq 80\%$, whereas Bob uses an APD with efficiency $\eta \approx 20\%$, dead time $100$ $\upmu$s, and gate window 33.5~ns at 15 MHz. The heterogeneous detectors result in widely varying link efficiencies and reflect the types of variability in quantum resources that should be expected in larger quantum networks. %

Due to random birefringence effects induced by the single-mode optical fiber,  careful compensation of polarization rotation must be performed in order to realize high-fidelity entanglement distribution. Given the fact all fibers are located either indoors or underground, we did not find it necessary to perform active polarization tracking~\cite{Xavier2008, Xavier2009} the polarization state was typically stable for hours at a time. However, we did perform manual compensation before each experiment utilizing FPCs at the outputs of the WSS. Following an alignment procedure similar to previous fiber-based polarization sources~\cite{Wang2009}, we first inserted a polarizer after the biphoton source that permitted only $H$-polarized photons to pass through, adjusting each FPC to minimize the counts when the analyzers are set to measure the $V$ state, i.e., fast axis of the QWP (HWP) oriented at 0$^\circ$ (45$^\circ$) with respect to horizontal. This ensures compensation of $H/V$ individually, up to an unknown phase between $H$ and $V$.

This residual phase amounts to an output state of the form $\ket{HV}+e^{i\phi}\ket{VH}$, which we can compensate by defining effective $D/A$ axes. That is, setting the fast axis of the QWP for both receivers to 45$^\circ$ with respect to horizontal, we tune the HWP angle on one of the two photons to maximize coincidences and define the $D/D$ measurement for both receivers: all other basis state projections are obtained by rotating the HWP by fixed amounts relative to each $D$ setting. This correction can be understood intuitively on the Poincar\'{e} sphere as a rotation that maintains an equal $H/V$ superposition but changes the relative phase between them~\cite{Peters2005b}. Specifically, defining QWP and HWP angle pair ($\theta_Q,\theta_H$), we have the measurement settings $H=(0^\circ,0^\circ)$, $V=(0^\circ,45^\circ)$, $D=(45^\circ,x^\circ)$, $A=(45^\circ,x^\circ+45^\circ)$, $R=(45^\circ,x^\circ+22.5^\circ)$, and $L=(45^\circ,x^\circ-22.5^\circ)$, where $x$ is a free parameter chosen to maximize contrast. In the current experiment, we optimize the $x$ of one polarization analyzer individually for each pair of nodes, but it is important to note that the number of free parameters (three) is sufficient to optimize the HWP settings for all pairs of nodes \emph{simultaneously}, by solving a system of equations or monitoring correlations in complementary bases in real-time, similar to the techniques mentioned in Refs.~\cite{Poppe2004, Wengerowsky2018}.

\subsection{Bandwidth Allocation 1}
\label{Bandwidth Allocation 1}
The eight-channel polarization-entangled source supports a variety of bandwidth allocations that can be tailored to match a desired network configuration. Through the WSS, adjustments to bandwidth provisioning can be made in real-time without modifying any fiber connections. This approach for entanglement distribution was recently realized in a tabletop experiment~\cite{Lingaraju2020a}, and here we demonstrate it in a deployed fiber-optical network for the first time.

Given the imbalance in system efficiency, due to both the deployed fiber loss and heterogeneous detector technology, we explore and test two different bandwidth allocations. A patch-panel to patch-panel link loss of 1.8~dB (3.3~dB) was measured from Alice to Bob (Charlie); combined with the detectors used, Charlie and Alice enjoy the highest overall efficiency, followed by Alice and Bob and finally Bob and Charlie. In the first allocation, we seek to balance the entanglement rates by assigning the spectral slice with lowest flux (Ch.~8) to Charlie and Alice (C--A), the brightest slice (Ch.~1) to Alice and Bob (A--B), and the remaining (Ch.~2--7) to Bob and Charlie (B--C).

\begin{figure*}[tb!]
	\centering
	\includegraphics[width=0.75\textwidth]{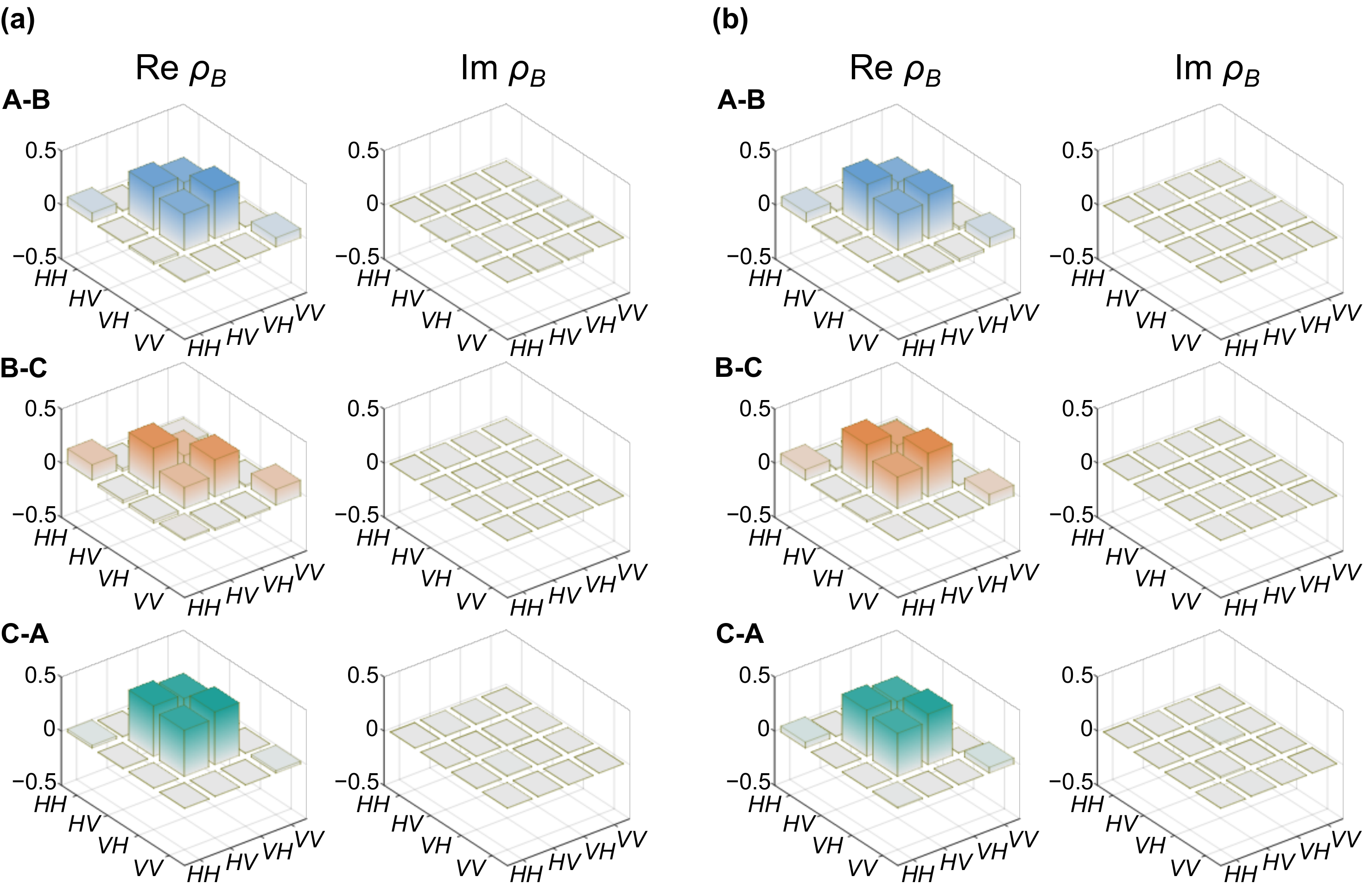}
	\caption{Density matrices estimated by polarization tomography for each pair of users for (a) Allocation 1 and (b) Allocation 2.}
	\label{alloc_density}
\end{figure*}

At a pump power of 15.6~mW, the average singles counts in this allocation for each pair (in units of s$^{-1}$) are: $1.2\times10^5$ and $5.0 \times 10^3$ (A--B), $7.5 \times 10^3$ and $1.4 \times 10^5$ (B--C), and $1.0 \times 10^4$ and $4.7 \times 10^4$ (C--A). %
To measure state quality, we perform polarization tomography on data obtained with a 10~ns coincidence window and 60~s integration time. Figure~\ref{alloc_density}(a) shows the Bayesian mean estimated~\cite{Lukens2020b} density matrices for each pair of nodes, again based on measurements in the $H/V$ and $D/A$ bases. In these tests, only the logical link under test is bright (the other channels are blocked) to reduce the background at each detector; this is equivalent to spectrally resolved detection at each node. Nevertheless, no accidentals are subtracted from these results, and the fidelities with respect to $\ket{\Psi^+}$ follow in Table~\ref{alloc}.

From fidelities alone, it is not immediately clear how useful the provided entanglement in the network is for generic quantum communications tasks. Since we have performed full state tomography, this could in principle be predicted for any desired protocol. But for greater generality, we propose an ``entanglement bandwidth'' analogous to bit rates in classical links. Distillable entanglement---the asymptotic rate of pure Bell pairs which could be produced from copies of the state, local operations, and classical communications~\cite{Bennett1996}---seems especially fitting as such a quantifier, but it is extremely difficult to compute, even with full knowledge of the density matrix. As an alternative, we consider the log-negativity $E_\mathcal{N}$~\cite{Vidal2002}, which provides an upper bound on distillable entanglement. (For two qubits specifically, $E_\mathcal{N}>0$ is also a necessary and sufficient condition for nonseparability~\cite{Horodecki1996}.) When multiplied by the coincidence rate, $E_\mathcal{N}$ thus gives a generic metric for link performance in terms of entangled bits (ebits) per second ($R_E$). Computing $E_\mathcal{N}$ and $R_E$ from the Bayesian samples, we obtain the results in the final two columns of Table~\ref{alloc}. Importantly, even the lowest-fidelity link (B--C) possesses an $R_E$ greater than zero by approximately three standard deviations.

\begin{table}[tb!]
	\centering
	\caption{Link data for both bandwidth allocations.}
	\label{alloc}
		\begin{ruledtabular}
	\begin{tabular}{c|c|c|ccc}
		\textbf{Alloc.} & \textbf{Link} & \textbf{Ch.} & \textbf{Fidelity} & $\bm{E_\mathcal{N}}$ \textbf{[ebits]} & $\bm{R_E}$ \textbf{[ebits/s]} 		\\ \hline
		\multirow{3}{*}{1} 
		& A--B  &  1 	&  $0.75 \pm 0.03$ & $0.70 \pm 0.08$ & $56  \pm 6$ 	\\
		& B--C  &  2--7 &  $0.55 \pm 0.06$ & $0.4 \pm 0.2$ & $30  \pm 10$ \\ 
		& C--A  &  8    &  $0.90 \pm 0.01$ & $0.89 \pm 0.03$ & $206 \pm 6$ 	\\ 	
		\hline
		\multirow{3}{*}{2} 
		& A--B  & 3     &  $0.75 \pm 0.03$ & $0.70 \pm 0.09$ & $57  \pm 7$ 	\\  
		& B--C  &  1--2 &  $0.69 \pm 0.04$ & $0.6 \pm 0.1$ & $26  \pm 4$ 	\\ 
		& C--A  &  4    &  $0.84 \pm 0.02$ & $0.82 \pm 0.05$ & $320 \pm 20$ \\
	\end{tabular}
	\end{ruledtabular}
\end{table}

\subsection{Bandwidth Allocation 2}
\label{Bandwidth Allocation 2}
In the second allocation, our goal was to improve the lowest fidelity in the network and test the performance of different allocations for the other links. In general, a tradeoff exists when  selecting channel allocations: assigning more channels to a pair of users increases the number of entangled photons,  boosting the coincidence rate, yet it can also result in lower fidelity due to increased accidentals (which scale quadratically with flux) and higher sensitivity to wavelength-dependent birefringence (from  the broader bandwidth). How these effects play out in the overall entanglement rate $R_E$ is difficult to predict. So to explore the interplay between them, we next assign Ch.~1--2 to B--C, Ch.~3 to A--B, Ch. 4 to C--A, leaving the remaining four channels unassigned as examples of resources available for future use should additional nodes come online.

The average singles counts in Allocation~2 (units of s$^{-1}$) are: $1.2\times10^5$ and $5.1 \times 10^3$ (A--B), $6.4 \times 10^3$ and $6.7 \times 10^4$ (B--C), and $1.9 \times 10^4$ and $8.4 \times 10^4$ (C--A). 
The density matrices obtained follow in Fig.~\ref{alloc_density}(b), along with the specific fidelities and entanglement rates in Table~\ref{alloc}. Compared with Allocation~1, the A--B link maintains similar performance in all categories. B--C experiences an appreciable increase in fidelity but, due to the reduced flux, a slightly lower $R_E$. Interestingly, the higher flux on the C--A link reveals the exact opposite behavior: lower fidelity but higher $R_E$. These examples highlight that the optimal allocation may depend on a given objective: what characteristics are most valuable to the users on a network? In light of the variety of possible answers to this question, bandwidth provisioning with the WSS offers exceptional flexibility compared to a fixed wavelength configuration: it can adapt and meet changing demands, apportioning resources for seamless service to each link without physical disconnections~\cite{Lingaraju2020a}.

\section{Remote State Preparation}
\label{Remote State Preparation}

As a proof-of-principle application of our QLAN, we performed the RSP protocol between network nodes.
RSP is a quantum communications protocol similar but simpler than teleportation.  Like teleportation, it requires users to share entanglement and classical communications. However, RSP uses a single-photon measurement on half the entangled pair instead of Bell state analysis on the to-be-teleported input photon and half the entangled pair.  In RSP, the sender measures one photon of an entangled pair in order to prepare the remaining particle at the receiver in some desired 
quantum state~\cite{Lo2000, Pati2000, Bennett2001, Peters2005a}. %
We note this is non-deterministic, though in some cases, the classical communications enable one to apply a unitary operation to yield the desired state, or to simply post-select the successful events as we have done here. Within the network layer architecture of Sec.~\ref{Network Architecture}, RSP resides in the application layer, leveraging the logical connections provided by the transport layer---i.e., the entanglement resources as summarized in Table~\ref{alloc}. The ideal Bell state $\ket{\Psi^+}$ is maximally correlated in all three standard polarization bases (rectilinear, diagonal, and circular): $\ket{\Psi^+}\propto \ket{HV} + \ket{VH} = \ket{DD} - \ket{AA} = \ket{RR} - \ket{LL}$. (We adopt the convention $\ket{R}\propto\ket{H}+i\ket{V}$ and $\ket{L}\propto\ket{H}-i\ket{V}$.) Thus, a simple RSP experiment follows from measuring in one of these bases and performing tomography on the remaining photon, comparing the result to the ideal case.

\begin{figure}[tb!]
	\centering
	\includegraphics[scale=0.38]{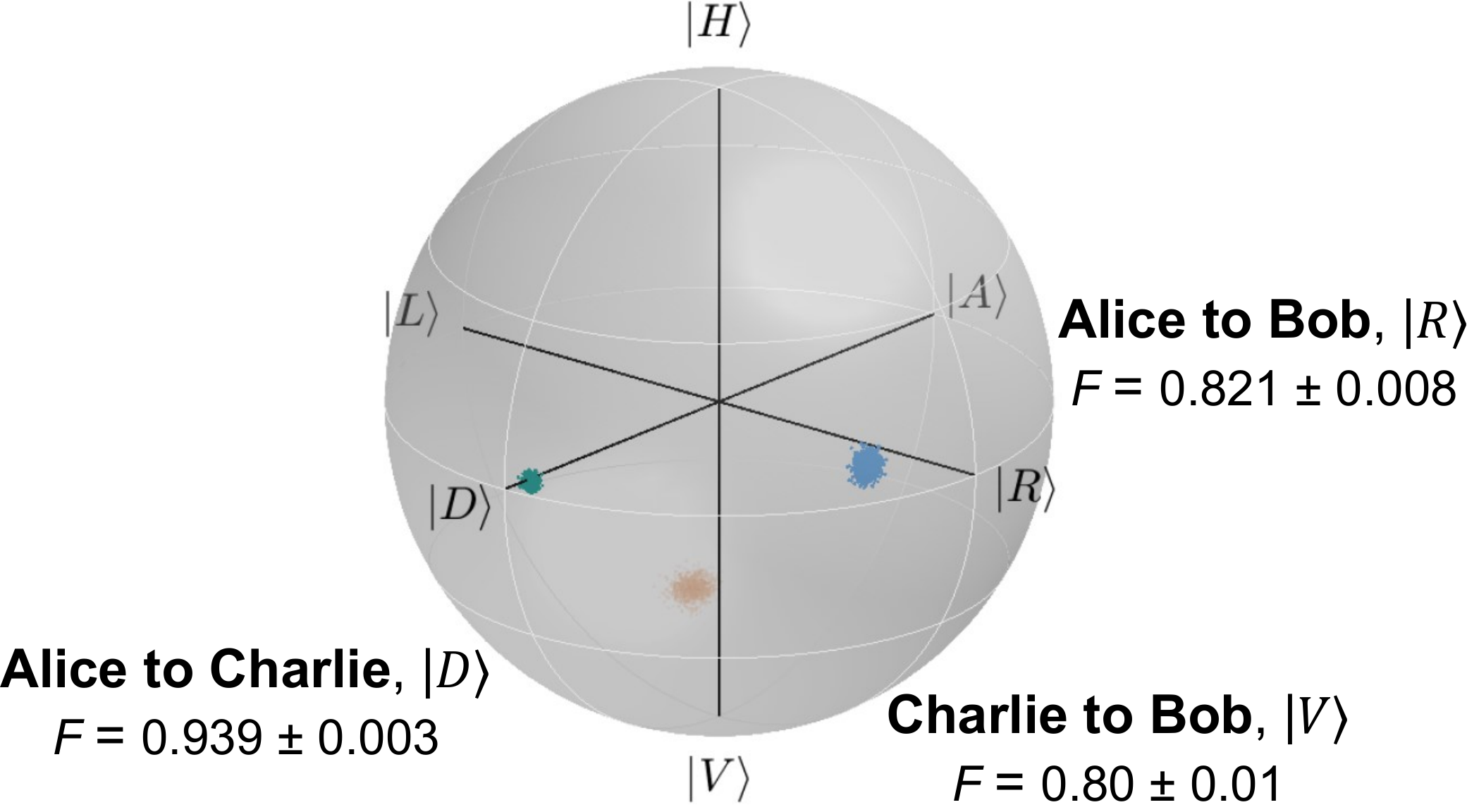}
	\caption{Tomography results for qubit states remotely prepared on each logical link.}
	\label{rsp_sphere}
\end{figure}

We implement RSP of a single state on each link of the network, utilizing Allocation~2 above. For link A--B, Alice prepares the state $\ket{R}$ at Bob by projecting her photon onto $\ket{R}$; for B--C, Charlie prepares the state $\ket{V}$ at Bob by projecting his photon onto $\ket{H}$; and for C--A, Alice prepares the state $\ket{D}$ at Charlie by projecting her photon onto $\ket{D}$. The results of Bayesian tomography on the prepared qubits are plotted on the Poincar\'{e} sphere in Fig.~\ref{rsp_sphere}, based on measurements in a complete set of two-dimensional mutually unbiased bases ($H/V$, $D/A$, and $R/L$). Each cloud consists of 1024 samples from the posterior distribution, giving a visual indication of the uncertainty in each result~\cite{Lu2020a}. Fidelities with respect to the target state are provided for each link, including error bars from the standard deviation of the retrieved samples. As expected, the C--A link gives the highest fidelity and lowest uncertainty, followed by A--B and B--C. The experimentally measured RSP states are in excellent agreement with predictions based on the density matrices in Fig.~\ref{alloc_density}(b): their fidelities with respect to the relevant partially traced density matrix are $\sim$0.99 in all cases, confirming that the nonidealities in Fig.~\ref{rsp_sphere} result from the underlying entanglement quality and not errors in the RSP protocol. To our knowledge, these results represent the first demonstration of RSP over a deployed network,  confirming the ability of our QLAN to provide the network stack needed to successfully carry out  a quantum communications protocol.

\section{Discussion}
\label{Discussion}
In this work, we have successfully demonstrated the first adaptive bandwidth provisioning for entanglement in a deployed network, showing how allocations impact both fidelity and entanglement rates and enable dynamic quality-of-service provisioning. %
All results above were presented without accidental subtraction, confirming that appreciable entanglement is distributed to all user pairs under current conditions. However, from a technical perspective, reducing noise from accidentals would offer major performance improvements. Table~\ref{accSub} displays the results obtained for both allocations if we do subtract accidentals prior to tomography, where the accidentals correspond to the coincidences in a 10~ns window that is time-shifted from the correlated peak. The fidelities in all cases are $>92\%$, suggesting that addressing such noise limitations should improve the quality of service significantly. (Intriguingly, the accidentals-subtracted entanglement rates $R_E$ are equal within error to the raw results in Table~\ref{alloc}. While we are unaware of any theoretical requirement for this relationship, intuitively it makes sense: subtracting accidentals eliminates noise from the data, but it does not increase the throughput of entangled pairs.)

Perhaps the most effective technical direction to reduce noise would be improving the time synchronization across the QLAN. With the current GPS clocks, our coincidence windows are 10~ns, yet the detector-jitter--limited coincidence peaks are much less than 1~ns wide. Accordingly, reducing network jitter---through an optical sync or White Rabbit---and using a 1~ns coincidence window would reduce accidentals 10-fold while maintaining the number of correlated detections at the present level. Under such conditions, the added noise from all channels transmitting simultaneously would prove much more manageable. %

In general, polarization entanglement seems to be particularly well-suited to QLANs, where relatively short and environmentally protected fibers result in low polarization mode dispersion and minimal drift with time. Indeed, with the level of stability obtained in submarine fibers, polarization encoding has been harnessed successfully on scales of $\sim$100~km without active compensation~\cite{Wengerowsky2019}. By replacing the current biphoton source with one covering the entire C-band (1530--1570 nm) or beyond~\cite{Herbauts2013,Vergyris2017,Zhu2019} and leveraging nested WSSs as in Fig.~\ref{fig2}(d), one can envision much larger networks utilizing the same technology. As an example, suppose that the bandwidth of the source in Fig.~\ref{fig2}(d) is divided into three sections, each frequency-correlated. %
One section is distributed among the outputs of the first WSS and the second section to those of the nested WSS (i.e., the first WSS passes this entire section to the second WSS). This enables each WSS to form a QLAN where nodes in the same QLAN are able to form a fully connected network. The third section can then be routed or divided between the WSSs (on-demand) to form logical links between these two QLANs (internetworking). In this way a basic quantum internet, in the literal sense as ``interconnected networks,'' can be realized with small upgrades to the present system. %

Finally, the question of how best to characterize the generic quality of a  quantum network remains open. As we have observed here, optimizing one metric may negatively impact others, and so the best configuration may depend strongly on the application. Interestingly, this goal of quantum network benchmarking mirrors similar objectives in quantum computing, where the concept of quantum volume has been proposed to account for both the number of qubits and gate performance in a consistent framework~\cite{Cross2019}. So while it remains to be seen what (if any) metric will prove the most informative for evaluating quantum networks, our exploration of an entanglement rate in ebits/s forms an important step in this direction, as it is able to integrate both the quality and quantity of entanglement resources into a single value.

\begin{table}[b!]
	\centering
	\caption{Accidentals-subtracted link data for both bandwidth allocations.}
	\label{accSub}
		\begin{ruledtabular}
	\begin{tabular}{c|c|c|ccc}
		\textbf{Alloc.} & \textbf{Link} & \textbf{Ch.} & \textbf{Fidelity} & $\bm{E_\mathcal{N}}$ \textbf{[ebits]} & $\bm{R_E}$ \textbf{[ebits/s]} 		\\ \hline
		\multirow{3}{*}{1} 
		& A--B  &  1 	&  $0.960 \pm 0.005$ & $0.96 \pm 0.01$ & $52.4  \pm 0.6$ 	\\
		& B--C  &  2--7 &  $0.925 \pm 0.006$ & $0.95 \pm 0.03$ & $33  \pm 1$ 	\\
		& C--A  &  8    &  $0.985 \pm 0.001$ & $0.987 \pm 0.005$ & $207  \pm 1$ 	\\
		\hline
		\multirow{3}{*}{2} 
		& A--B  & 3     &  $0.957 \pm 0.004$ & $0.97 \pm 0.02$ & $53.3  \pm 0.8$ 	\\ 
		& B--C  &  1--2 &  $0.959 \pm 0.005$ & $0.98 \pm 0.01$ & $24.8  \pm 0.4$ 	\\
		& C--A  &  4    &  $0.968 \pm 0.001$ & $0.98 \pm 0.01$ & $320  \pm 4$ 	\\
	\end{tabular}
	\end{ruledtabular}
\end{table}

\section*{Acknowledgments}
We thank S. Hicks for assistance in setting up the control plane. This work was performed in part at Oak Ridge National Laboratory, operated by UT-Battelle for the U.S. Department of Energy under contract no. DE-AC05-00OR22725. Funding was provided by the U.S. Department of Energy, Office of Science, Office of Advanced Scientific Computing Research, through the Early Career Research Program and Transparent Optical Quantum Networks for Distributed Science Program (Field Work Proposals ERKJ353 and ERKJ355). N.B.L. acknowledges funding from the Quantum Information Science and Engineering Network (QISE-NET) through the National Science Foundation (1747426-DMR). B.J.L. and Y.Y.P. were sponsored by the U.S. Department of Energy, Office of Science, Basic Energy Sciences, Materials Sciences and Engineering Division (Field Work Proposal ERKCK51). C.E.M. was sponsored by the Intelligence Community Postdoctoral Research Fellowship Program at the Oak Ridge National Laboratory, administered by Oak Ridge Institute for Science and Education through an interagency agreement between the U.S. Department of Energy and the Office of the Director of National Intelligence.

\end{document}